\documentclass{article}

\usepackage{arxiv}

\usepackage[utf8]{inputenc} 
\usepackage[T1]{fontenc}    
\usepackage{hyperref}       
\usepackage{url}            
\usepackage{booktabs}       
\usepackage{amsfonts}       
\usepackage{nicefrac}       
\usepackage{microtype}
\usepackage{amsmath}
\usepackage{cleveref}       
\usepackage{lipsum}         
\usepackage{graphicx}
\usepackage{natbib}
\usepackage{doi}

\title{Decentralized Finance and Local Public Goods: A Bayesian Maximum Entropy Model of School District Spending in the U.S.}

\newif\ifuniqueAffiliation
\uniqueAffiliationtrue

\ifuniqueAffiliation 
\author{ {\hspace{1mm}Juan Melo}\\
	Ecole d'économie de la Sorbonne - UFR 02\\
	University Paris 1 Panthéon-Sorbonne\\
    Paris, FR. \\
	\texttt{jm5557@columbia.edu} \\
	}
\else
\usepackage{authblk}

\newbox{\orcid}\sbox{\orcid}{\includegraphics[scale=0.06]{orcid.pdf}} 
\author[1]{%
	Juan Melo\thanks{\texttt{jm5557@columbia.edu}}}%

\affil[1]{Ecole d'économie de la Sorbonne - UFR 02, Pittsburgh, University Paris 1 Panthéon-Sorbonne, Paris FR}

\fi

\renewcommand{\headeright}{}
\renewcommand{\undertitle}{}
\renewcommand{\shorttitle}{A Bayesian Maximum Entropy Model of School District Spending in the U.S.}

\hypersetup{
pdftitle={Decentralized Finance and Local Public Goods: A Bayesian Maximum Entropy Model of School District Spending in the U.S.},
pdfsubject={q-bio.NC, q-bio.QM},
pdfauthor={Juan Melo},
pdfkeywords={Public School Expenditures, Statistical Equilibrium, Quantal Response Statistical Equilibrium (QRSE), Bayesian Inference, Markov Chain Monte Carlo (MCMC) Sampling, Tiebout Hypothesis, Inter-jurisdictional Competition, Decentralized Education Funding, Socioeconomic Disparities, Fiscal Policy Analysis.},
}

\begin{document}
\maketitle

\begin{abstract}
This paper investigates the distribution of public school expenditures across U.S. school districts using a bayesian maximum entropy model.  Covering the period 2000-2016, I explore how inter-jurisdictional competition and household choice influence spending patterns within the public education sector, providing a novel empirical treatment of the Tiebout hypothesis within a statistical equilibrium framework.  The analysis reveals that these expenditures are characterized by sharply peaked and positively skewed distributions, suggesting significant socio-economic stratification. Employing Bayesian inference and Markov Chain Monte Carlo (MCMC) sampling, I fit these patterns into a statistical equilibrium model to elucidate the roles of competition, as well as household mobility and arbitrage in shaping the distribution of educational spending. The analysis reveals how the scale parameters associated with competition and household choice critically shape the equilibrium outcomes. The model and analysis offer a statistical basis for shaping policy measures intended to affect distributional outcomes in scenarios characterized by the decentralized provision of local public goods.
\end{abstract}


\section{Introduction}
\label{sec-1}

\subsection{Tiebout Competition}
\label{sec-1-1}

A central problem in the economic analysis of the provision of local
public goods is the lack of incentives of voters to reveal their true
demand. \citet{tiebout_pure_1956} proposed to study the problem of
local public goods through a quasi-market model in which
consumer-voters express their preferences for local public goods by
moving in and out of local jurisdictions.The Tiebout hypothesis states
that local jurisdictions will tend to sort into homogenous blocks with
respect to demand for local public goods and tax levels, when these
are taken to be a form of prices in the model. The core idea behind
this hypothesis is that a Tiebout sorting equilibrium, if it exists,
will eliminate inefficiencies associated with demand diversity;
households will not be forced to pay higher tax levels than they would
otherwise prefer, nor are they are able to free-ride on neighboring
households' relatively higher contributions to the local tax-service
package.

Tiebout's 1956 paper was, and continues to be, an important catalyst
for renewed research in the analysis of decentralized government
finance. Tiebout's major contribution was to challenge the standard
belief of the time that there was no market-based solution to the
problem of local public goods provision. He did so by placing
geographic location and mobility at the core of the analysis, and by
using the latter as a proxy for choice and preference revelation.

\begin{quote}
There is no way in which the consumer can avoid revealing his
preferences in a spatial economy. Spatial mobility provides the
local-goods counterpart to the private market's shopping trip.
Tiebout (1956)
\end{quote}

While \citet{tiebout_pure_1956} agreed with
\citet{musgrave_voluntary_1939} and \citet{samuelson_pure_1954}, that
the determination of federal government expenditures could only have a
political solution, he argued for explaining variations in local
government expenditures in terms of decentralized sorting mechanisms
(a market analogy) and not by alluding to simple majority voting
schemes.

Tiebout's hypothesized mechanism of competition, when seen though the
lens of neoclassical equilibrium models, may be understood as having
three essential traits \citep{nechyba_tiebout_2020}. The first is
that when local communities are viewed as analogous to competing
firms, decentralization will allow for the optimal provision of public
services in the presence of heterogeneous household demands. The
second is the notion that competition will reduce incentives for local
governments to behave like `Leviathans' \citep{jha}. The latter
notion rests on the belief that the decentralized procurement and
provision of local public goods will counter the tendency of
governments to arbitrarily extract higher taxes from their residents
\citep{brennang,jimenez_is_2010}. A third feature implies that in
'equilibrium', the Tiebout mechanism will lead households to sort (to
some degree) on the basis of ability-to-pay and household income. This
latter feature is, of course, far from being unequivocally
desirable. The characterization of Tiebout sorting as an optimal
outcome, possessing intrinsic merit mostly on account of its capacity
to bring about productive efficiencies, turns out to be at odds with
basic legal notions regarding citizens' rights to education
\citep{jha}. An equilibrium in which public school expenditures
and quality are highly correlated with household characteristics
presents non-negligible moral and legal challenges. The scope of these
challenges has been duly evidenced by the continued legal battles and
policy debates over funding inequities in the US public education
system for the past 50 years
\citep{bakere,hertert_school_1994}. The fact that the optimal
outcome in a highly idealized formulation of the Tiebout hypothesis
turns out to be fundamentally at odds with what may be desirable at
the policy or household level (or is at the very least highly
contestable), does not rule out the possibility that sorting and the
rationing of government resources are in fact shaped by Tiebout-like
forces. It does however pose serious challenges to the modeling and
specification of the microeconomic primitives which drive the
competitive process.

One of the fundamental problems that comes out of the use of applied
general equilibrium models is that they force us to consider observed
economic distributions as resulting chiefly and mechanically from the
interaction of optimizing agents (households and governments) whose
preferences are fully satisfied. This is a modeling strategy that
rules out a priori the possibility that agents' expectations will
remain unfulfilled in equilibrium.

In the context of the economic analysis of the determinants of
heterogeneity in public school expenditure levels and demand, where
much of the theoretical and policy debates center around the explicit
recognition that education markets are structured by complex political
and production processes, the requirement that fully optimizing
behavior be consistent with observed equilibria is hard to
sustain. Furthermore, in the absence of plausible characterizations
for the microeconomic and political environments, it is hard to see
how any useful insights may be extracted from the study of general
equilibrium forces and outcomes.

This concern has steered the Tiebout and education finance literature
towards a path of building models of increased mathematical and
computational complexity, where elements such as heterogenous voting
preferences and non-financial inputs are incorporated in order to
provide richer descriptions that are more empirically relevant, as
well as plausible from a microeconomic perspective
\citep{nechyba_school_2003, tim}.

There has been a recent shift in the literature from building general
equilibrium models to building computational equilibrium models that
straddle a wide spectrum covering both purely theoretical and
empirically motivated formulations.  As \citet{nechyba_tiebout_2020}
notes, all such models start by explicitly specifying the underlying
mathematical structure of the economic environment being modeled. That
is, they provide a fully structural specification for household
preferences, school production functions, distributions for household
characteristics in the model (such as income), as well as mathematical
descriptions for the political process (voting models), the fiscal
environment, and the housing and private school markets. Through
simulation studies, the study of the equilibrium outcomes in these
models is then expected to yield meaningful policy insights, and to
provide a sandbox for experimenting with out-of-sample policy
interventions.

The problem is that the relevance of these simulation
studies hinges on the empirical plausibility of the elaborate
microeconomic structure that is being used to represent the underlying
mechanics of the data generating process, and on the confidence we may
have in the model's parameters to adequately capture empirically
relevant processes. But if we consider the fact that the task of
determining the empirical plausibility of any given model
specification for complex social environments with large degrees of
freedom may be ill-posed and underdetermined \citep{schafol}, then
it is hard to see how the route of increasing model complexity in
fully micro-founded general (or computational) equilibrium models is
likely to yield unambiguous and normatively unbiased results. There is
a very broad continuum of models and solutions that are consistent
with any set of circumstantial data and evidence
\citep{golan_foundations}. And misspecification can show up either
at the level of functional forms (production and preference
functions), criterion or decision functions, the specification of
voting models, as well as priors for stochastic inputs in the model
(e.g. household characteristics).

This paper takes an alternative approach that makes use of maximum
entropy methods and a statistical equilibrium framework to model and
study the effect of competition in shaping the distributions of local
government education expenditures for the period of 2000-2016 in the
United States. The advantages of this maximum entropy/statistical
equilibrium framework are plenty, but a central one that we consider
here is that it allows us to study the competitive dynamics of the US
public education market (a complex social systems with large degrees
of freedom) without having to commit \emph{a priori} to a heavy
mathematical scaffolding of the underlying microeconomic
environment. Rather, it allows is to study one plausible way in which
the probabilistic structure of school district expenditures can be
seen to emerge from a pair of parsimonious behavioral and institutional
constraints that we place on the underlying microeconomic environment.

\subsection{Unintended Outcomes, Market Efficiency and Tiebout Sorting}
\label{sec-1-2}

One of the interesting aspects of Tiebout's original 1956 formulation
is that it remains non-committal with respect to any specific
equilibrium model formulation, even if it highlights a set of stylized
facts and features that the hypothetical competitive process should
meet. But as the history of the empirical tests of the Tiebout
hypothesis has shown
\citep{nechyba_tiebout_2020,oates_effects_1969,edel_taxes_1974},
it is not truly possible to test all of the assumptions of the larger
Tiebout hypothesis at once without running into contradictions. For
example, testing the assumption of residential mobility alongside the
capitalization of fiscal variables into housing prices may run against
Tiebout's larger efficiency hypothesis (since the presence of
capitalization is evidence for the existence of excess demand for
housing in the jurisdiction where taxes and local service levels are
being capitalized) \citep{epple_fiscal_2004}>. Similarly, as we
pointed above, the existence of Tiebout sorting is to be better
understood as a potentially unexpected macroeconomic outcome (to at
least some section of households).  Seen under this light, the
prospect of being able to reconcile the underlying political
contradictions of the education market with the assumption of fully
maximizing households in a general equilibrium model seems
far-fetched. That said, we believe there is need and ample room to
focus on some aspects of the Tiebout hypothesis, and that it is
possible to study the empirical support for the general claim that
expenditures in local public goods are heavily shaped (and at least
partially explained) by competitive forces and a boundedly rational
arbitrage that takes place at the household level in terms of
education consumption.

This paper applies the theoretical framework of the Quantal Response
Statistical Equilibrium (QRSE) model developed in \citet{schafol}. As
mentioned above, the approach taken by the paper is not fully agnostic
with respect to microeconomic structure, as it utilizes an entropy
constrained model of residential mobility and jurisdictional choice as
the baseline characterization of household behavior. This baseline
model makes the behavioral assumption that households try to maximize
the rate of return on tax expenditures (considered as prices for local
education services), under the constraint of a limited capacity to
process market and political signals. In the context of low-income and
inner-city households, we put forward the idea that this limited
capacity may also be interpreted as a form of restricted economic
agency. The basic outline of this behavioral model is very similar to
the one found in Sims' rational inattention program
\citep{sims_implications_2003}.  Through the inclusion
of an information-theoretic constraint on the utility maximizing program of
households, this baseline specification delivers a meaningful
probabilistic description of household behavior.

\subsection{Sample and Paper Structure}
\label{sec-1-3}

The statistical equilibrium distribution of the QRSE model presented
here is a positively skewed unimodal distribution of the household
rate of return for local tax expenditures, with four parameters $T$,
$S$, $\mu$, and $\alpha$ that qualitatively predict the observed
data and give insights into the possible range of variation across
sub-sampling schemes. We use US public education finance data for all school districts
in the period of 2000-2016. We then apply Bayesian inference and MCMC sampling to fit the observed
distribution for the entire period to the theoretical QRSE model, and to recover posterior
distributions for the four unknown parameters.

The paper consists of 5 sections. Section 2 provides a description of
the data used and presents empirical the frequency distributions for the
key fiscal and expenditure variables used in building the
model. Section 3 puts our application of the QRSE model into context
by discussing fiscal decentralization, Smith's theory of
competition, and the measurable implications of the Tiebout
hypothesis. Section 4 then develops the paper's QRSE treatment of
Tiebout competition and derives the statistical equilibrium density
for the \emph{local per pupil rate of return on tax and service charges},
which we term \emph{educational returns}.  Section 5 describes the
Bayesian estimation of the model, and discusses results for the four
main parameter estimates.

\section{School District Variables}
\label{sec-2}

\subsection{Data}
\label{sec-2-1}

This paper uses data from the National Center for Education
Statistics’ Common Core of Data, the US Census Bureau Small Area
Income and Poverty Estimates, and the US Department of Education’s
EDFacts initiative \footnote{This data has been made available in a
harmonized format in a publicly available API by the Urban Institute,
which provides a convenient and reliable interface to all the major
federal dataset. Education Data Portal (Version 0.10.0), Urban
Institute, accessed February,
2021,\url{https://educationdata.urban.org/documentation/}, made available
under the ODC Attribution License.}.

We consider a sample of local expenditures in primary and secondary
education, local taxes, enrollment and population estimates for all 50
US states and school districts (on average \textasciitilde{} 13,500), in the 2000-2016
period.  We excluded a total of 233 data points (roughly 0.1\% of the
dataset), 58 of which were due to extreme value observations
attributable to data entry error, and the remaining 175 due to missing
values in one of the outcome variables. The total number of
observations for all 50 US states and school districts (on average
13,500) is $N= 229,553$. The outcome variable we are seeking to characterize is defined as:

\begin{equation}
\label{xdef}
x = \frac{\mbox{Total Local Education Expenditures}}{Enrollment} -
\frac{\mbox{Total Local Taxes and Charges}}{\mbox{Population}}
\end{equation}

The outcome variable $x$ is the \emph{household per pupil rate of return on tax and
service charges}, which we term \emph{educational returns}.  The variable \emph{Total Local Education Expenditures} is
aggregated from a large set of expenditure categories in primary and
secondary education (K-12) that include \emph{instruction}, \emph{textbooks},
\emph{pupil support services}, \emph{staff} , \emph{transportation},
\emph{administration}, \emph{maintenance}, \emph{food services}, \emph{utilities},
\emph{supplies}, and \emph{technology}. We denote the variable \emph{Total Local
Education Expenditures}, scaled by enrollment, as $\kappa$.

The variable \emph{Total Local Taxes and Charges} aggregates the following
revenue categories: \emph{Private contributions}, \emph{fines and forfeits},
\emph{property sales}, \emph{rents and royalties}, \emph{sales and services},
\emph{individual and corporate income taxes}, \emph{general fees}, \emph{public
utility taxes}, \emph{general sales taxes}, \emph{and property taxes}. We denote
the variable \emph{Total Local Education Expenditures}, scaled by school
district population, as $\tau$. 

\medskip




Due to constraints from missing data or comparability across regions
and years, this paper works with aggregate local revenue categories,
without excluding \emph{general fees} or \emph{service charges}.

\subsection{Empirical Distributions}
\label{sec-2-2}

\begin{figure}[htb]
\centering
\includegraphics[width=5in]{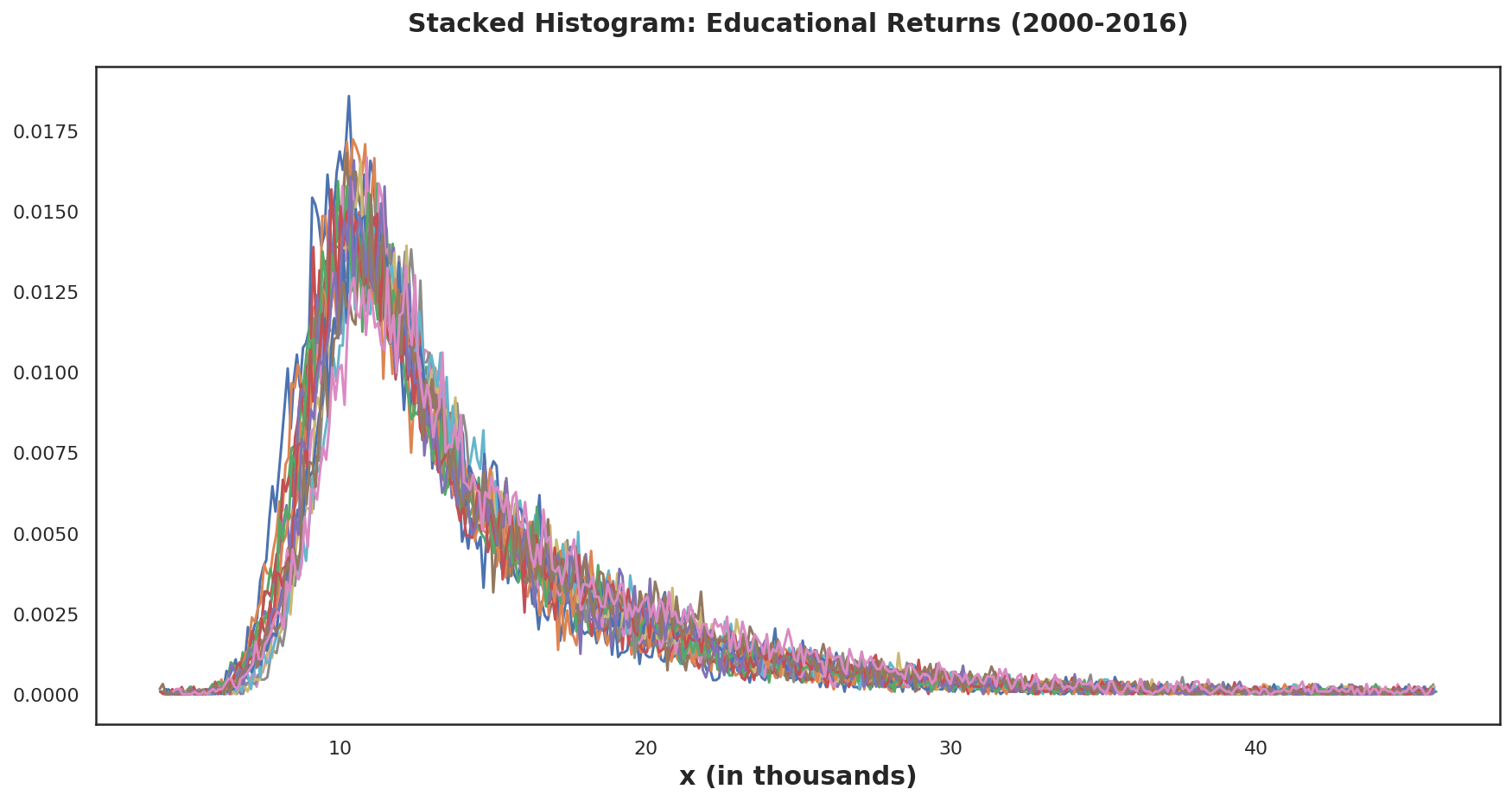}
\caption{\label{tbstack}Stacked histogram plots of the educational returns variable $x$ (in thousands). 2000-2016. The stacked histograms reveal the time invariance of the statistical equilibrium distributions for the different years that make up the pooled sample.}
\end{figure}

Below we present the marginal empirical distribution for $x$, as
defined in formula \ref{xdef}, for the period $2000-2016$. In figure
\ref{tbstack} we plot a stacked histogram with the empirical density
for each year in the pooled sample.  The stacked histogram reveals the
persistent organization of educational returns into highly peaked
asymmetric distributions with positive skew. The pattern variance in
the right tails is particularly revealing of disequilibrium
fluctuations in the Tiebout sorting process. Fatter right tails with
positive skew, we believe, might constitute a strong signal of
inter-jurisdictional sorting in the Tiebout sense.

\begin{figure}[htb]
\centering
\includegraphics[width=6in]{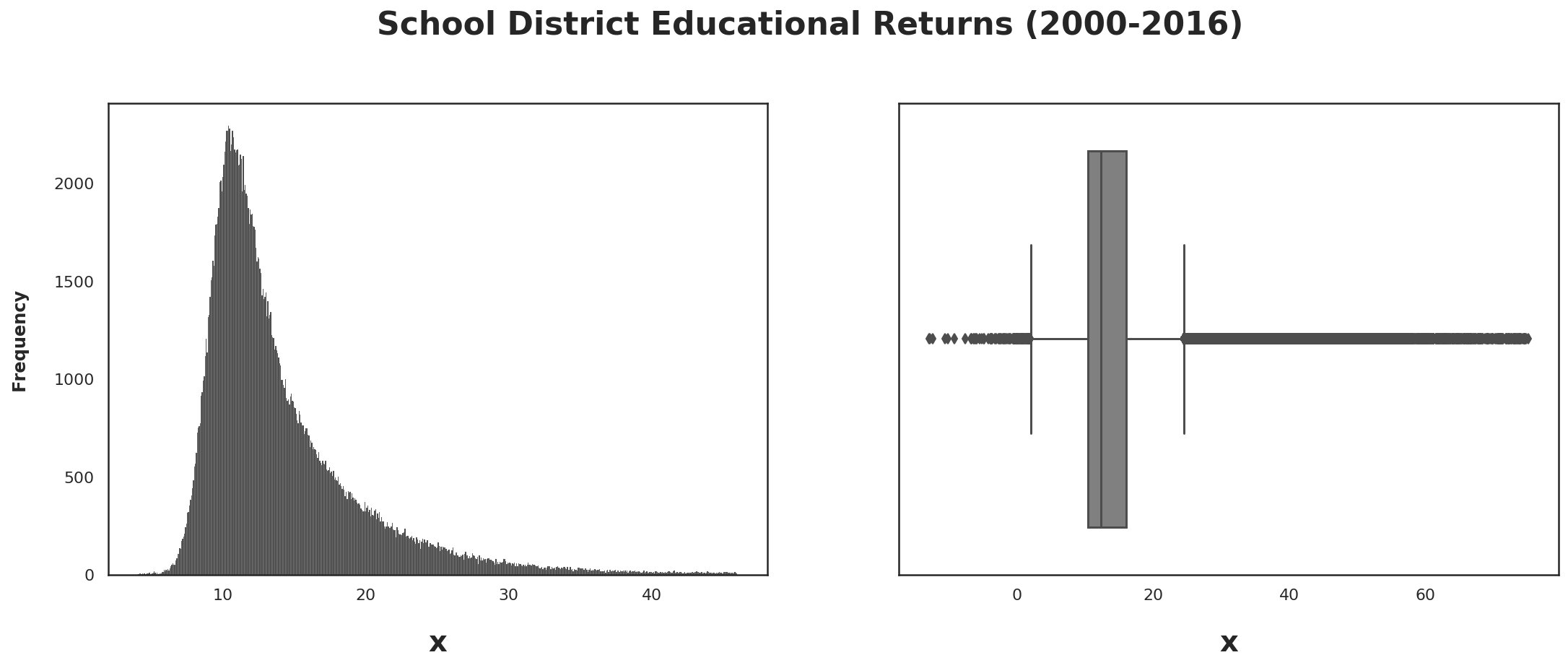}
\caption{\label{mdistx}Marginal distribution of x (in thousands) for the period $2000-2016$. Histogram and Box-Plot.}
\end{figure}

A visual inspection of these distributions points to the asymmetric
Subbotin or exponential power distribution \citep{alfarano_statistical_2012} as a potential candidate for modeling the statistical
equilibrium density and for approximating the empirical frequencies of
school district educational returns in the period considered
\citep{fruhwirth-schnatter_mixture_2015}>.

While these may be good candidates for characterizing highly skewed and
peaked distributions, in general, their specifications would not allows us to draw
straightforward theoretical conclusions from estimates of their
location, scale and shape parameters. In sections 4 and 5, we show how
the maximum entropy derivation of the QRSE model leads to a marginal
density function $\hat{f}_{x}$ whose parameter estimates can be
directly linked to the impact of competition and households incentives
on jurisdictional sorting and expenditure levels.

\section{Free Competition and The Tiebout Hypothesis}
\label{sec-3}

\subsection{Smith's Theory of Competition}
\label{sec-3-1}

In the classical Smithian theory of competition, profit-seeking agents
make the choice to enter or exit lines of production based on the
market's prevailing rate of return. Unlike general equilibrium models,
where prices and rates of return are understood as static market
clearing quantities in a pure exchange economy, competitive price and
rate discovery in the Smithian conception follows a process governed
by negative feedbacks. As producers enter profitable lines of
production they tend to lower the profit rate by crowding the output
and supply of that particular good. Eventually, this forces relocation
of capital and resources into other sectors, pushing the rate back to
attractive levels
\citep{smith_wealth_1937,schafol,shaikh_capitalism_2016}.

This homeostatic process, in which rates are pushed up and down as
resources enter and relocate throughout sectors in the economy, has
the particular advantage that it lends itself well to a probabilistic
interpretation \citep{farjoun_laws_2020}. Observed prices and
returns can be effectively seen as gravitating around a fundamental
central tendency. If we think in terms of probability distributions,
we may suitably express a rate's theoretical fundamental value or its
'natural' (regulating) level as the distribution's location
parameter. Similarly, the extent to which observed rates vary and the
intensity with which they respond to exit and entry decisions may be
suitably captured by variance and scale parameters.

In this paper we seek to link this theory of free competition to
Tiebout's original account of the role that inter-jurisdictional
competition plays in determining equilibrium levels for local fiscal
and expenditure variables, as well as for household choice.  Our
proposition departs radically from the education finance literature in
that we propose a statistical understanding of equilibrium, and do not
follow the requirement that parameter estimates be interpreted as
operating at fully efficient margins. This leaves open the possibility
that households' expectations and preferences might remain unfulfilled
in equilibrium. But because the equilibrium is statistical, and not
static, this does not constitute a barrier for analysis. The framework
considered here allows us to study how such departures from optimality
and efficiency may relate to the distinct statistical features of
expenditure levels in decentralized education markets, such as
positive skew and sharp pre-modal decay. 

\subsection{The Tiebout Hypothesis}
\label{sec-3-2}

In his 1956 paper, Tiebout laid out a set of highly abstract
assumptions for his model of local public goods competition, which he
also called "a pure theory of local expenditures". The model's
assumptions may be summarized as follows:

\begin{enumerate}
\item Consumer-voters have full mobility and knowledge of prevailing
expenditure patterns in neighboring communities.  \item Mobility is a
proxy for consumer-voter choice.  \item There is a large number of
communities from which to choose.  \item There is an optimal community
size, given demand conditions and fixed resources.  \item Tax-service
packages are set according to consumer-voter preferences.
\end{enumerate}

Given that Tiebout's model postulates consumer-voters as choosing
tax-service packages by moving in and out of communities, and that the
levels of these public goods packages are determined by local
governments in response to demand (i.e migration inflows and
outflows), we can see how the Smithian framework applies. In the
context of education expenditures, we assume that inhabitants are
looking for high rates of return to their decision to locate or
relocate to a particular community. We take these rates of return to
be proportional to the difference between the \emph{per capita local tax
rate} and \emph{per pupil local expenditures} that consumer-voters face in
the local public goods market.  For the purposes of this paper, we
take these market units to be school districts. Competitive school
districts will offer attractive per pupil expenditure rates, and low
per capita tax rates and service charges.  As consumer-voters crowd
districts with good schools and low taxes, the rates of educational
returns return will adjust accordingly, and under the assumptions of
full mobility and rational incentives to fulfill expectations in a
local public goods payoff, the iterative process of rate adjustment
and migration flows will stabilize expenditures into the observed
patterns.

The outcome variable $x$, the \emph{per pupil rate of return on local tax
spending and charges}, which we termed \textbf{educational returns} is
defined by the difference:

\medskip

\[ \mathbf{x = \kappa - \tau } \]

\medskip

where $\kappa$ is the total local expenditure per pupil, scaled by the
school district's enrollment, and $\tau$ the total local tax and
charge burden, scaled by the district's population. In the next
section we delve deeper into our statistical treatment of Tiebout
competition and derive the QRSE density $\hat{f}_{x}$.

\section{Local Public School Expenditures and Household Choice}
\label{sec-4}

\subsection{Statistical Equilibrium Modeling and Maximum Entropy Inference}
\label{sec-4-1}

The highly peaked and positively skewed patterns of the outcome
variable $x$ for the 2000-2016 period suggests the existence of a
central tendency in the distribution along with non-symmetric
deviations from its mean. Asymmetric Exponential Power Distributions
(AEPD) and Skewed Exponential Power Distributions (SEPD)
\citep{alfarano_statistical_2012,schafsim,mundt_asymmetric_2019}
are good candidates to model this kind of data. But because we need a
constructive probabilistic description that is phenomenologically
relevant, as well as theoretically interpretable in its parameters, we
implement a Quantal Response Statistical Equilibrium
(QRSE) model to fit this data.

The notion of statistical equilibrium has been widely used in physics
and information theory
\citep{jaynes_concentration_1983,jaynes_information_1957}. A
statistical equilibrium for a quantity $x$ takes the form of a
probability density function $f_{x}$; it represents the most likely
distribution for the outcome variable given a set of theoretical and
empirical conditions. It can be derived by maximizing the entropy
$H[f_{x}] = - \sum \limits_{x} f_{x} \, log[f_{x}]$ subject to
constraints expressing relevant information, theory or
observations. The methodology is most commonly used in the context of
bayesian statistics with the purpose of deriving informative priors by
feeding moment constraints and relevant background information to the
maximum entropy program. An important feature of the maximum entropy
program is that as long as the set of constraints provided describe a
non-empty convex set in the space of distributions, the maximum
entropy program will yield an optimal solution that can be used as the
statistical equilibrium density of the model
\citep{schafol,golan_foundations}. For more details on the
derivation of maximum entropy distributions see
\citep{sivia_data_2006,jaynes_probability_2003,golan_foundations}.

We can view single-state solutions to general equilibrium models as
special cases of this statistical equilibrium model. They represent
degenerate probability densities for the variable $x$ where only the
optimal solution is assigned a positive probability. Such degenerate
distributions, with all the probability mass concentrated around a
single point, also imply systems that operate with zero
entropy. 

Formally, the model we present here is a derived maximum
entropy distribution for the joint density of household jurisdictional
choice and educational returns $x$. Rather than giving full
statistical content to the complete set of assumptions in the Tiebout
hypothesis, we use this derived probability model to examine Tiebout's
intuition regarding the role of competition and household
choice in shaping the marginal distribution $f_{x}$.

\subsection{A Logit Quantal Response Function for Household Choice}
\label{sec-4-2}

The general Quantal Response Statistical Equilibrium (QRSE) model
presented here links a set of household quantal actions $a \in
\mathcal{A}$ to the outcome variable $x \in \mathbb{R}$. This could
also be a vector $\vec{x}$ in $\mathbb{R}^{n}$, but in this paper the
variable $x$ is a scalar, which corresponds to the level of
educational returns at the school district level. $\mathcal{A}$ is be
the binary action set $\mathcal{A} = \{e, s \}$ —where $e$ stands for the
\emph{entry} of households into a particular school district, and $s$ for
the \emph{exit}. The interaction between the hidden quantal action set
$\mathcal{A}$ and the outcome variable $x$ is modeled by the joint
distribution $f_{x,a}$.  The maximum entropy distribution $f_{x,a}$
represents a statistical equilibrium where the inflow/outflow actions
of households, represented by the set $\mathcal{A}$, are conditionally
dependent on the educational returns rate $x$, but also shape it via
equilibrating forces and the negative feedback process which we
defined as Smithian and Tiebout-like competition. We define the payoff for the typical household by the function 

\medskip

\begin{equation}\label{payoff} 
\pi(a, x): \mathcal{A}\times \mathcal{X} \rightarrow \mathbb{R}  
\end{equation}

\medskip

The payoff takes as input an action from the action set $\mathcal{A}$,
and a signal from the state space of educational returns $x \in
\mathcal{X}$.  We use linear symmetric payoffs such that $\pi(e,x) =
-\pi(s,x)$, as shown in equation \ref{pstruct}. 

\medskip 

\begin{equation} \label{pstruct}
\begin{aligned}
\pi(e,x) = x -\mu \\
\pi(s,x) = \mu - x \\
\end{aligned}
\end{equation}

\medskip 

The difference of the entry and exit payoffs is given by equation
\ref{paydiff} below:

\begin{equation} \label{paydiff}
\begin{aligned}
\Delta \mathrm{\pi}(a, x) &= \\
&= \pi(e,x)  - \pi(s,x) \\
&= 2(x-\mu) 
\end{aligned}
\end{equation}

\medskip 

This payoff structure contains a location parameter $\mu$ to express
the fact that households will have a tipping point for moving in or
out of a particular school district. Households will tend to move into
districts where the level of educational returns is above this
expectation $\mu$ and vice versa.  Note that $\mu$ is not the average
rate, but the expectation that the households forms prior to relocation.

The first constraint that we impose on our statistical model of local
education returns is that it be micro-founded by a probabilistic
theory of behavior. In other words, we expect the entry and exit
decisions of households to be non-deterministic responses to
variations in local expenditure patterns for the set of communities
that constitute the local public goods market. It is possible to think
of this as the assumption that households follow `mixed strategies' in
determining whether to move in or out of a particular district. At
times they will follow their payoff maximizing action, but sometimes
they won't. We expect the probabilities of observing a particular
behavior to be proportional to the payoffs in equation \ref{pstruct},
and exclude the degenerate case in which households choose only the
payoff maximizing action with probability $1$.

One way to derive the stochastic function which describes the
micro-level behavioral component of the model is to impose a minimum
entropy constraint on the utility maximization program of the
agent. The household payoff maximization program and the associated
Lagrangian take the forms shown below in equations \ref{maxent1} and
\ref{lag1}.

\medskip

\begin{equation} \label{maxent1}
\begin{aligned}
&\underset{f_{a| x}\geq 0}{\max} \space \sum_{\mathcal{A}} f_{a \mid x} \space
\pi(a,x) \\ &\text { s.t: } \\ & \sum_{\mathcal{A}} f_{a \mid x}=1
\\ &-\sum_{\mathcal{A}} f_{a \mid x} \log [f_{a \mid x}] \geq \space
H_{\operatorname{min}} \end{aligned} 
\end{equation}

\begin{equation} \label{lag1}
\begin{aligned}
\mathcal{L}=-& \sum_{\mathcal{A}} f_{a \mid x} \space
\pi(a,x)-\lambda\left(\sum_{\mathcal{A}} f_{a \mid x}-1\right)
\\ &+T\left(\sum_{\mathcal{A}} f_{a \mid x} \log [f_{a \mid x}]-H_{\min }\right)
\end{aligned} 
\end{equation}

\medskip

\medskip 

This maximization program introduces the behavioral parameter of the
model $T$. In \citet{schafol} it is described as a `behavior
temperature' parameter in analogy to statistical models of
thermodynamic systems, but it can also be understood as a bounded
rationality constraint.  Maximizing the payoff subject to a minimum
entropy constraint is dual to the problem of maximizing the entropy of
the mixed strategy $f_{a|x}$ subject to a minimum payoff
constraint. In that dual case, the Lagrangian containts the term
$\beta\left(\sum_{a} f_{a \mid x} \space
\pi(a,x)-U_{\operatorname{min}}\right)$, which links the multiplier
$\beta=\frac{1}{T}$ to the minimum payoff constraint.  

The solution to this programming problem yields a general logit
quantal response or Gibbs density, as in equation \ref{gibbs_micro}:

\begin{equation} \label{gibbs_micro}
f_{a \mid x}=\frac{e^{\frac{\pi(a, x)}{T}}}{\sum_{\mathcal{A}} e^{\frac{\pi(a, x)}{T}}}
\end{equation}

\medskip 

For the case of the binary action set $\mathcal{A}$, the program
yields the canonical QRSE logit quantal response functions in \ref{qr1}
and \ref{qr2}. 

\medskip

\begin{equation} \label{qr1}
\begin{aligned}
f_{e \mid x} &= \frac{1}{1+e^{-\frac{\Delta \mathrm{\pi}(a, x)}{T}}}  = 
\frac{1}{1+e^{-\frac{2(x-\mu)}{T}}}
\end{aligned}
\end{equation}

\begin{equation} \label{qr2}
\begin{aligned}
f_{s \mid x}=1-f_{e \mid x} = \frac{1}{1+e^{\frac{2(x-\mu)}{T}}}
\end{aligned}
\end{equation}

\medskip

This pair of stochastic quantal response functions take the shape of
the cumulative distribution function for the logistic
distribution. The parameter $T$ is the scale parameter that expresses
the sensitivity of the household choice rule to the difference in the
observed outcome from the subjective expectation ($x-\mu$). The
introduction of $\mu$ allows us to model household behavior as
'chasing' a central tendency in the outcome variable $x$, and as
having a `tipping point' for the choice to enter or exit a particular
school district. In the context of Tiebout competition, these
stochastic choice rules should be conceived as representing
conditional probabilities for migration inflow or outflow into the
ensemble of school districts for which the parameter $T$ is
estimated. They are not `agent-level' functions that additively
aggregate to the ensemble equilibrium distribution, but rather a
\emph{meso-level} description that models the dependency of entry/exit
flows on observed expenditure patterns and local fiscal variables.

\medskip

\begin{figure}[htb]
\centering
\includegraphics[width=6in]{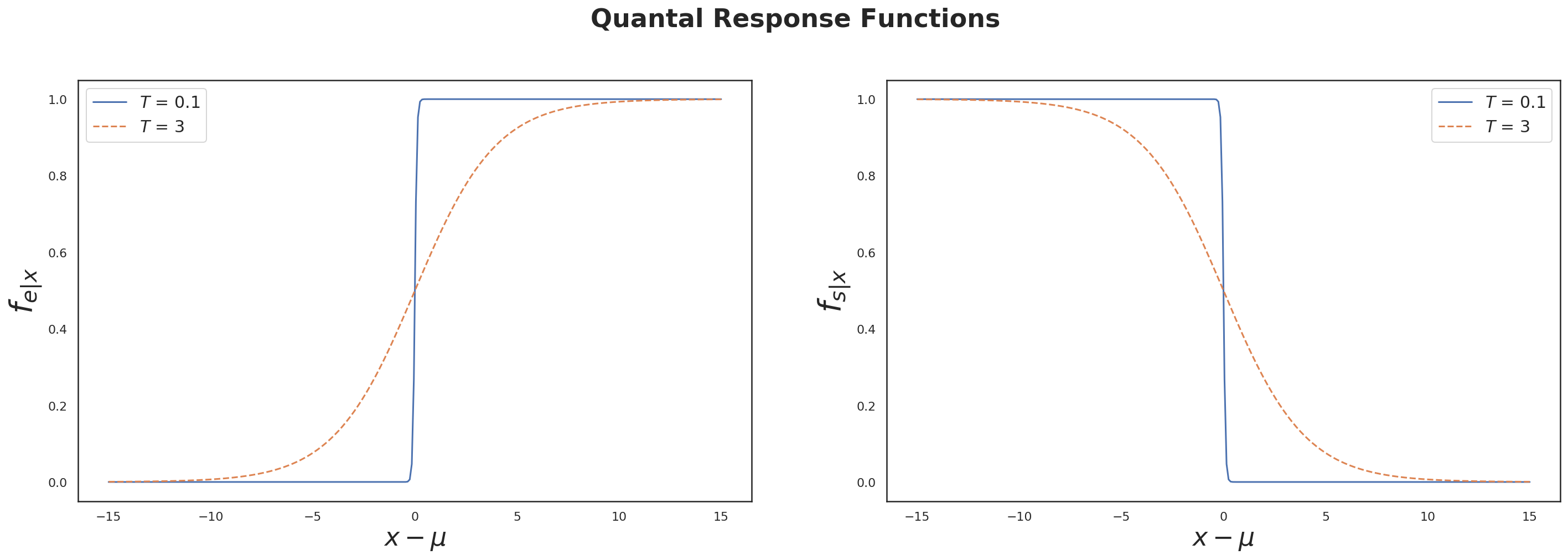}
\caption{\label{qexample}Quantal responses functions for probability of household entry/exit. At near zero $T$ the quantal response function takes resembles the heaviside step function. Under a zero entropy assumption $(T=0)$, we would expect households to move into districts exclusively in the case where $x - \mu  > 0$.  This graph is illustrative (taking $T$=0.1 and $T$ = 3 as examples).}
\end{figure}

\subsection{The Competitive Feedback Constraint}
\label{sec-4-3}

In the canonical QRSE model from \citet{schafol} which is applied here,
agents respond to payoff differentials by entering or exiting a
particular economic sector. In the Tiebout setting, household inflows
into particular districts may cause congestion in public school
services and lead to a feedback effect where the level of educational
returns $x$ will be reduced. Conversely, household outflows from
school districts will tend (over time) to push the educational returns
rate back up. In the QRSE model, this assumption takes the form of a
congestion or `competitive feedback' constraint on the outcome
distribution. The competitive feedback constraint is expressed as an
inequality that limits the scale of the difference between expected market
rates conditional on entry and exit, as shown in \ref{fbconst}. 

\begin{equation} \label{fbconst}
0 \leq f_{e} \: \mathrm{E}[(x-\alpha) |e ]-f_{s}\:\mathrm{E}[(x-\alpha) | s] \leq \epsilon\\
\end{equation}

\medskip 

This inequality expresses the idea that the expected jurisdictional
market rates will be higher conditional on entry than on exit——but
that their difference is small enough so that we wouldn't expect an
infinite inflow or crowding into a particular district. This
constraint allows us to parsimoniously model the simultaneous and
feedback driven relationship that exists between household choice and
expenditure levels in the local education market.

The mean outcome level $\bar{x}$ is then co-determined by a complex (non-reductive)
interaction between those two layers of the economic
process. In practice we tend to find that $\alpha \neq \mu$, which means that
the market sustains unexpected outcomes and unfulfilled expectations,
an assumption that seems appropriate for the case of public goods markets. The constraint in \ref{fbconst} can be unpacked in more detail using
the form in equation \ref{fbconst2}, where it is written as an
expectation of the market outcome $(x-\alpha)$, factored by the
difference in mixed strategy probabilities $\Delta \; f_{a|x}$:

\begin{equation} \label{fbconst2} 
\begin{aligned}
&f_{e}\int f_{x|e}\:(x-\alpha) \: dx - f_{s}\int f_{x|s} \: (x-\alpha) \: dx\\
&\\
&= \int f_{e|x}\:f_{x}\;(x-\alpha) dx - \int f_{s|x}\:f_{x}\;(x-\alpha) dx\\
&\\
&= \int \Delta f_{a|x} \: f_{x} \; (x-\alpha) \; dx 
&\\
&=  \int \operatorname{tanh}\left(\frac{x-\mu}{T}\right) f_{x} \; (x-\alpha) \space \; dx \\
& \leq \epsilon \\
&\\
\end{aligned}
\end{equation}

\medskip

The $\operatorname{tanh}$ function arises from the definition of the
logit quantal response functions, as shown below in \ref{tanhd}:

\begin{equation} \label{tanhd}
\begin{aligned}
& \Delta f_{a|x} = f_{e|x} - f_{s|x} \\
&\\
&=\left(\frac{1}{1+e^{-\frac{2(x-\mu)}{T}}}\right) - \left(\frac{1}{1+e^{\frac{2(x-\mu)}{T}}}\right)
&\\
&= \frac{e^{2(x-\mu)/T}-1}{e^{2(x-\mu)/T}+1}
&\\
&=\operatorname{tanh}\left( \frac{2(x-\mu)}{2T}\right)
&\\
&=\operatorname{tanh}\left( \frac{x-\mu}{T}\right)
&\\
\end{aligned}
\end{equation}

\medskip 

Thus, the competitive feedback constraint can be written using the general form written in
\ref{fbconst3}, noting that $\Delta f_{a|x} = \operatorname{tanh}\left(
\frac{x-\mu}{T}\right)$: 

\begin{equation} \label{fbconst3}
\int \Delta f_{a|x} \; f_{x} \; (x-\alpha) \; dx \; \leq \; \epsilon
\end{equation}

\medskip 

In the next subsection we explain how the assumptions of entropy constrained behavior and the
existence of a competitive feedback constraint determine, via the maximum
entropy program, the joint distribution $f_{a,x}$.

\subsection{QRSE Maximum Entropy Program and Density}
\label{sec-4-4}

The maximum entropy program for the QRSE model constrains the joint
distribution $f_{a,x}$ so that it is consistent with the following two
propositions:

\medskip

\begin{enumerate}
\item The behavioral property of a non-zero entropy rule for household
jurisdictional choice (entry/exit decisions).  \item The competitive
feedback constraint that we postulate for the Tiebout-like process in
analogy with the Smithian theory of competition.  \end{enumerate}

\medskip

These conditions were formally defined in the previous two
subsections. We also constraint the distribution so that it meets the
usual normalization condition: $\int f_{x} dx=1$. Hence, the
program maximizes the joint entropy of $f_{a,x}$ subject to
normalization, competitive feedback, and bounded household
choice constraints. We can express the joint entropy $\operatorname{H}_{x.a}$
in terms of the marginal entropy $\operatorname{H}_{x}$ and the conditional or
'binary entropy'  $\operatorname{H}_{a|x}$  \citep{cover2006elements},
as shown in \ref{jointh} and \ref{binaryh}.

\medskip

\begin{equation} \label{jointh}
\operatorname{H}_{x.a} = \operatorname{H}_{x}  + \int\limits_{\mathcal{X}}f_{x} \; \operatorname{H}_{a|x} \; dx
\end{equation}

\begin{equation} \label{binaryh}
\operatorname{H}_{a|x}= - \sum\limits_{\mathcal{A}}\; f_{a|x} \; \operatorname{log} [f_{a|x}]
\end{equation}

\medskip

Using the above decomposition, we can write the
maximization program for the QRSE model using the compact form shown
below in equation \ref{maxent2}.

\medskip

\begin{equation} \label{maxent2}
\begin{aligned}
&\max_{f_{x} \geq 0} \; \operatorname{H}_{x}  + \int\limits_{\mathcal{X}}f_{x} \; \operatorname{H}_{a|x} \; dx\\ 
&\text { st. } \\ 
& \int f_{x} \; dx \; =1 \\
&\int \Delta f_{a|x} \; f_{x} \; (x-\alpha) \; dx \; \leq \; \epsilon
\end{aligned}
\end{equation}

\medskip

Note that the second constraint parsimoniously encodes both the
behavioral and the market feedback constraints. The associated
Lagrangian takes the form in equation \ref{lag2} below:

\medskip

\begin{equation} \label{lag2}
\begin{aligned}
\mathcal{L}[f_{x}, \lambda, \gamma]=& \:\operatorname{H}_{x}  + \int\limits_{\mathcal{X}}f_{x} \; \operatorname{H}_{a|x} \; dx \\
&\\
&-\lambda\left(\int f_{x} d x-1\right) \\
&\\
&-\gamma\left(\int \Delta f_{a|x} \; f_{x} \; (x-\alpha) \; dx \; - \; \epsilon\right)
\end{aligned}
\end{equation}

\medskip

The multiplier associated to the competitive feedback constraint in the program
yields the $\gamma$ parameter for the candidate statistical
equilibrium density, which measures the effect of the competitive
feedback process on the marginal distribution $f_{x}$. 

The solution to this maximum entropy program produces a predictive
density $\hat{f}_{x}$ that is consistent with our 
description of Tiebout-like competition in local public goods
markets. The distribution $\hat{f}_{x}$ predicts the marginal
frequencies of the outcome variable $x$ and completes the theory by
determining the conditional densities $f_{x|e}$ and $f_{x|s}$, the
joint densitiy $f_{x,a}$, a well as the expectations $E[x|s]$ and
$E[x|e]$.

\medskip

The solution $\hat{f}_{x}$ takes the form of a Gibbs/Boltzmann
distribution, shown below in equation \ref{qkernel0}:

\medskip

\begin{equation} \label{qkernel0}
\hat{f}_{x}= \frac{e^{\operatorname{H_{a|x}}}\;e^{-\gamma\left(\Delta f_{a|x}\right)\;(x-\alpha)}}{\mathrm{Z}}
\end{equation}

\medskip

where $\mathrm{Z}$ is the partition function.

\medskip

\begin{equation} \label{partfunc}
\mathrm{Z} = \int \limits_{\mathcal{X}} e^{\operatorname{H_{a|x}}}\;e^{-\gamma\left(\Delta f_{a|x}\right)\;(x-\alpha)}\; \operatorname{dx}
\end{equation}

\medskip

By expressing the $\gamma$ parameter as $\gamma = \frac{1}{S}$, we can
rewrite the predictive marginal density $\hat{f}_{x}$ as below in
\ref{qkernel}:

\begin{equation} \label{qkernel}
\hat{f}_{x}= \frac{e^{\operatorname{H_{a|x}}}\;e^{-\operatorname{tanh}\left( \frac{x-\mu}{T}\right)\;\left(\frac{x-\alpha}{S}\right)}}{\mathrm{Z}}
\end{equation}

\medskip 

\medskip 

With this parametrization it is then possible to perform inference
using two scale parameters $T$ and $S$, and two location parameters
$\mu$ and $\alpha$, which all have the same dimension as the
educational returns variable $x$. The scale parameter $S$ accounts for
the concentration of educational returns around the mode that arises from the
market level process of jurisdictional competition, while the scale
parameter $T$ accounts for the concentration of values that arises
from the purposive behavior of households.

\medskip 

In the next section we provide details on the bayesian estimation of
the model for the pooled dataset using all US school districts in the
period 2000-2016, and focus our discussion on theoretically
interpretable results for the four unknown parameters $T$, $\mu$, $S$
and $\alpha$.

\section{Bayesian Estimation of QRSE Model}
\label{sec-5}

\subsection{MAPs and Distance Measures}
\label{sec-5-1}

We use Bayesian inference to recover the values for the unknown
parameter vector $\Gamma = [T, S, \mu, \alpha]$, for the full sample
containing all US school districts in the 2000-2016 period. The
approach we followed in our estimation procedure was to first find
close to optimal values for the model by jointly minimizing the
Kullback-Leibler divergence ($D_{KL}$) between the observed marginal
frequency $\bar{f}_{x}$ and the inferred theoretical frequency
$\hat{f}_{x}$. This is equivalent to finding maximum a posteriori
(MAP) point estimates for $\Gamma$, given that maximizing the
likelihood turns out to be equivalent to minimizing the KL-Divergence
(See \citet{golan_foundations}) . To find these MAPs we used the
available optimization packages found in the Python Scipy library
(COBYLA and SLSQP). Minimizing the functional in equation \ref{klfunc}
yields the MAPs that we then use as starting points for the MCMC
sampler.

\medskip 

\begin{equation} \label{klfunc} D_{K L}\left(\hat{f}_{x} \|
\bar{f}_{x} \right)=\sum \hat{f}_{\Gamma;x} \log
\left[\frac{\hat{f}_{\Gamma;x}}{\bar{f}_{x}}\right] \end{equation}

\medskip 

Additionally, we use the Soofi information distinguishability
statistic (Soofi ID; See \citet{soofi_information_2002} for details and
theory) to evaluate fit performance. The Soofi ID is shown below in
equation \ref{soofiq}. Smaller values of the KL-Divergence and of the
Soofi ID imply better model fits and the Soofi ID in particular gives
a measure of how much informational content is explained (recovered)
by the candidate distribution.

\medskip 

\begin{equation} \label{soofiq} I D\left(\hat{f}_{x}: \bar{f}_{x}
\right)=1-\exp \left[-D_{K L}\left(\hat{f}_{x} \| \bar{f}_{x}
\right)\right] \end{equation}

\medskip

\subsection{Model Specification and Markov-Chain Monte Carlo Sampling}
\label{sec-5-2}

We use the QRSE density itself as the likelihood for estimation,
considering that the sampler holds the data $D$ fixed as it explores
different probabilities for the parameters in $\Gamma$ via
$P(D|\Gamma)$. Alternatively, to justify this, one might simply note
that the likelihood is proportional to the sampling distribution ;
$L(\Gamma \mid x) \propto f_{x \mid \Gamma}$. The QRSE log-likelihood used for the sampler is shown below in
\ref{loglike}:

\begin{equation} \label{loglike} \operatorname{log}[\hat{f}_{x}] = \;
\operatorname{H_{a|x}} -\; \operatorname{tanh}\left(
\frac{x-\mu}{T}\right)\;\left(\frac{x-\alpha}{S}\right) -
\operatorname{log}\left(\mathrm{Z}\right) \end{equation}

\medskip

We directly compute the partition function $\mathrm{Z}$ by the sum in \ref{partsum}. 

\begin{equation} \label{partsum}
\mathrm{Z} = \sum \limits_{\mathcal{X}} e^{\operatorname{H_{a|x}}}\;e^{-\operatorname{tanh}\left( \frac{x-\mu}{T}\right)\;\left(\frac{x-\alpha}{S}\right)}
\end{equation}

\medskip 

We evaluate the log-likelihood in \ref{loglike} by computing sequences
of random samples from the joint posterior distribution of
$\Gamma$. In this paper we use a standard Metropolis-Hastings
algorithm (MCMC-MH; see \citet{hogg2018data}).  Our code uses PyMC3 \citep{salvatier_probabilistic_2015}, an open source probabilistic programming
framework written in Python \footnote{Code used, data and MCMC
sample traces will be made available in a
public GitHub repository for review. For details on the PyMC3 library
see: \url{https://docs.pymc.io/api/inference.html}}. 

\medskip 

For each parameter, we run 3 chains with $30,000$ iterations and
$4,000$ tuning samples. All of the chains converged with $\hat{R}
=1$. For more details on the convergence statistic $\hat{R}$ used see
\citet{vehtari2019rank}. We show a plot of the chain sample
traces below in figure \ref{mctrace}. In figure \ref{pairplot} we show
pair plots of the posterior samples for the four parameters, which do
not appear to be correlated. We used truncated normal priors centered
near the MAP estimates for T and S, with lower and upper bounds at
$0.1$ and $8$ respectively. For $\mu$ and $\alpha$ we used normal
priors centered near the MAPs and specified large variances in order
to explore reasonably wide ranges of the parameter space. Given
knowledge about the plausible ranges for the scale and location
parameters, along with the MAP estimates, this choice of weakly
informative priors seemed appropriate.

\begin{figure}[htb]
\centering
\includegraphics[width=5.5in]{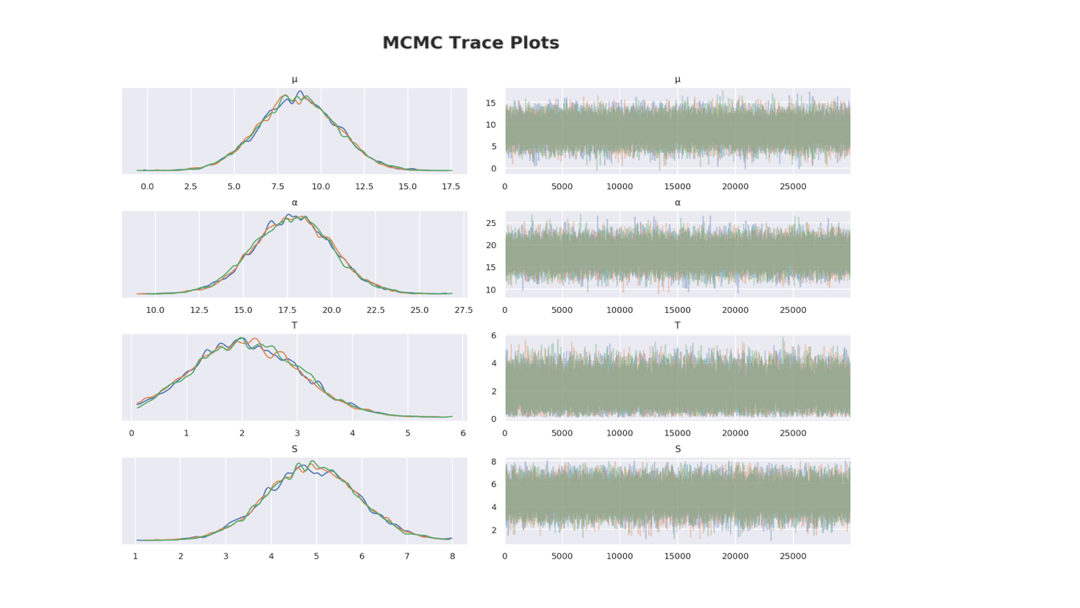}
\caption{\label{mctrace}Trace plots for MCMC samples obtained using the Metropolis-Hastings algorithm. 3 chains, 30,000 iterations per chain and 4,000 tuning samples.  All US School Districts. 2000-2016.}
\end{figure}

\begin{figure}[htb]
\centering
\includegraphics[width=5.5in]{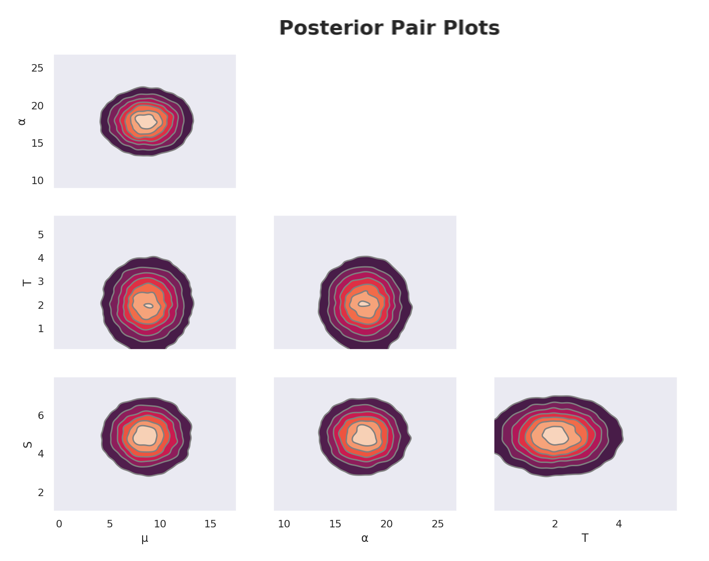}
\caption{\label{pairplot}Posterior pair plots for parameters T, S, $\mu$ and $\alpha$.}
\end{figure}

\clearpage

\subsection{Results}
\label{sec-5-3}

\begin{figure}[htb]
\centering
\includegraphics[width=5in]{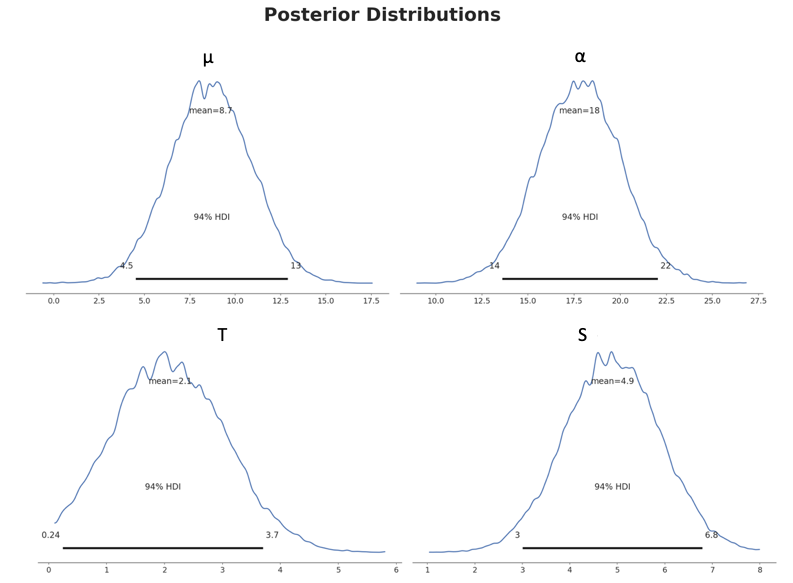}
\caption{\label{posteriors}Posterior distributions for T, S, $\mu$, and $\alpha$.All US School Districts. 2000-2016.}
\end{figure}

Table \ref{tabpost} gives the summary statistics for the estimated
parameters T, S, $\mu$, and $\alpha$. In figure \ref{posteriors} we
plot the four posterior distributions for the QRSE parameters, which
are unimodal, symmetric and have relatively wide standard deviations.

\begin{table}[htbp] 
 \centering
  \begin{tabular}{l l l l l}  
  \multicolumn{5}{c}{ \textbf{\textbf{Posterior Estimates Summary}} }\\ \\ \hline 
      Parameter & Mean (Sd)  & Mode & 94 \% HDI & $\hat{R}$ \\ \hline
       &              &              &        &    \\
                $\mu$   &  8.66 (2.24)  &   7.78 &    [4.54, 12.9] &    1.0 \\
                $\alpha$   &  17.8 (2.24)  &  19.71 &  [13.61, 22.05] &    1.0 \\
                T   &   2.1 (0.94)  &   2.17 &     [0.24, 3.7] &    1.0 \\
                S   &   4.9 (1.01)  &   4.69 &    [3.01, 6.79] &    1.0 \\  \\\hline
  \end{tabular} 
  \caption{\label{tabpost} Summary statistics of estimated parameters T, S, $\mu$, and $\alpha$. The means, standard deviations, 94\% credible intervals and the convergence statistics $\hat{R}$ from the MCMC samples are reported. All US School Districts, 2000-2016. } \label{poststats}
\end{table}

\clearpage

\subsection{Fits}
\label{sec-5-4}

\begin{figure}[htb]
\centering
\includegraphics[width=6in]{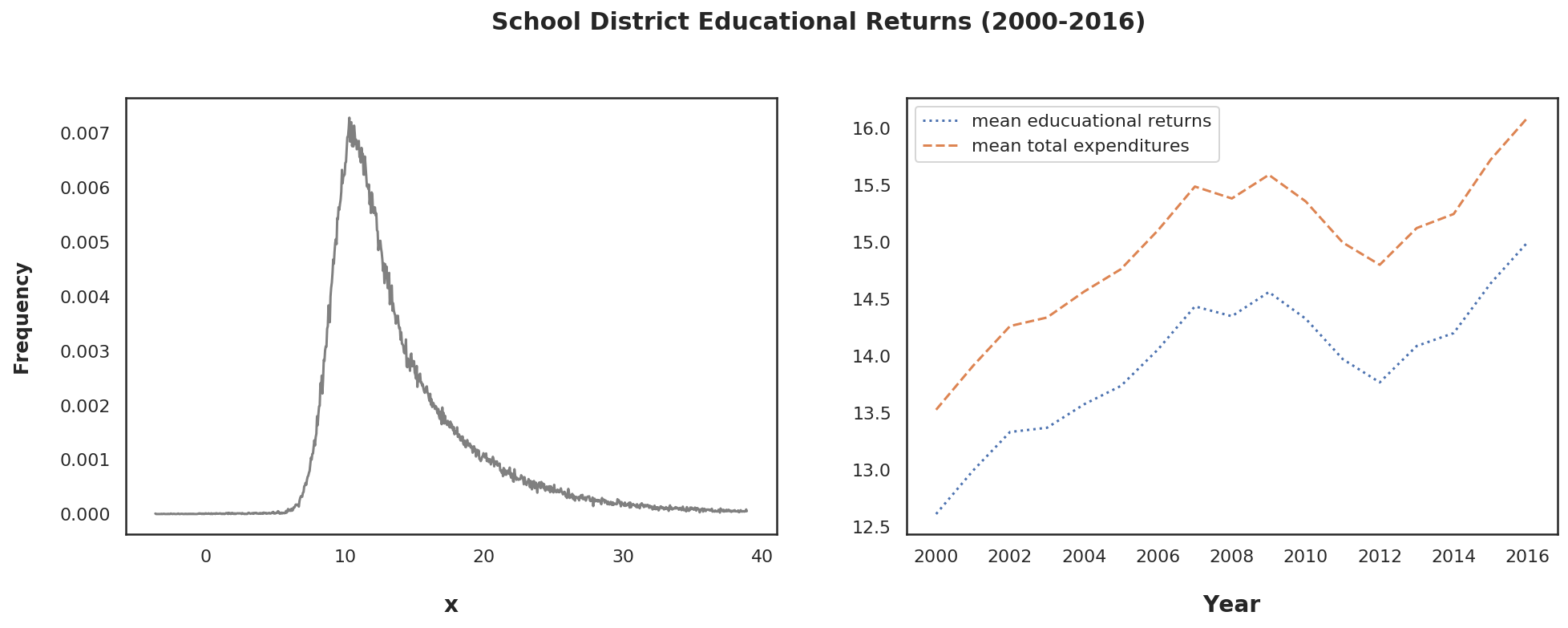}
\caption{\label{mdistx2}Marginal distribution of x, and yearly line plot for $x$ and $\kappa$ for the period 2000-2016. The histogram excludes the upper $0.01$ quantile (for visualization). Plotted in thousands. All US School Districts. 2000-2016.}
\end{figure}

In table table \ref{sumstats} we present summary statistics for the fiscal
variables used in the model, and for the educational returns variable
$x$. In  figure \ref{mdistx2} we plot the line histogram of $x$ for the entire
2000-2016 period, alongside yearly time series for both mean
educational returns $x$ and mean total expenditures $\tau$.

\begin{table}[htbp] 
 \centering
  \begin{tabular}{l l l l l}  
  \multicolumn{5}{c}{ \textbf{\textbf{Summary Statistics}} }\\ \\ \hline 
      Variable & Mean  & S.D. & Min. & Max.  \\ \hline
      Educational Returns ($x$) & 14.27 &  6.30 & -12.83 &   75.08\\ 
      Total District Expenditures ($\kappa$) & 15.31 &  6.69 &   0.01 &  102.56\\ 
      Taxes and Charges ($\tau$) & 1.04 &  1.04 &   0.00 &   59.55  \\ \hline
  \end{tabular}
  \caption{\label{sumstats}Model variables (in thousands). All US School Districts, 2000-2016. } 
\end{table}

\begin{figure}[htb]
\centering
\includegraphics[width=5in]{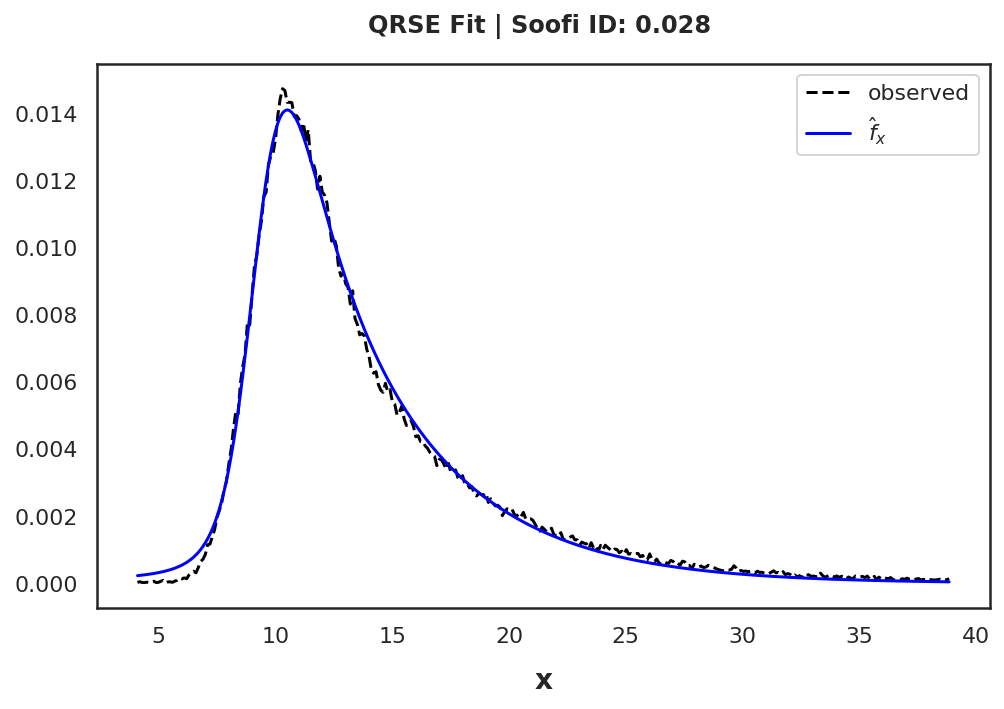}
\caption{\label{qrsefit}Line histogram of observed distribution for $x$ (educational returns). Overlaid is the fitted predictive marginal density $\hat{f}_{x}$. We excluded the upper $0.01$ quantile (for visualization). The Soofi ID/performance fit measure is shown. All US School Districts. 2000-2016.}
\end{figure}

\begin{figure}[htb]
\centering
\includegraphics[width=6in]{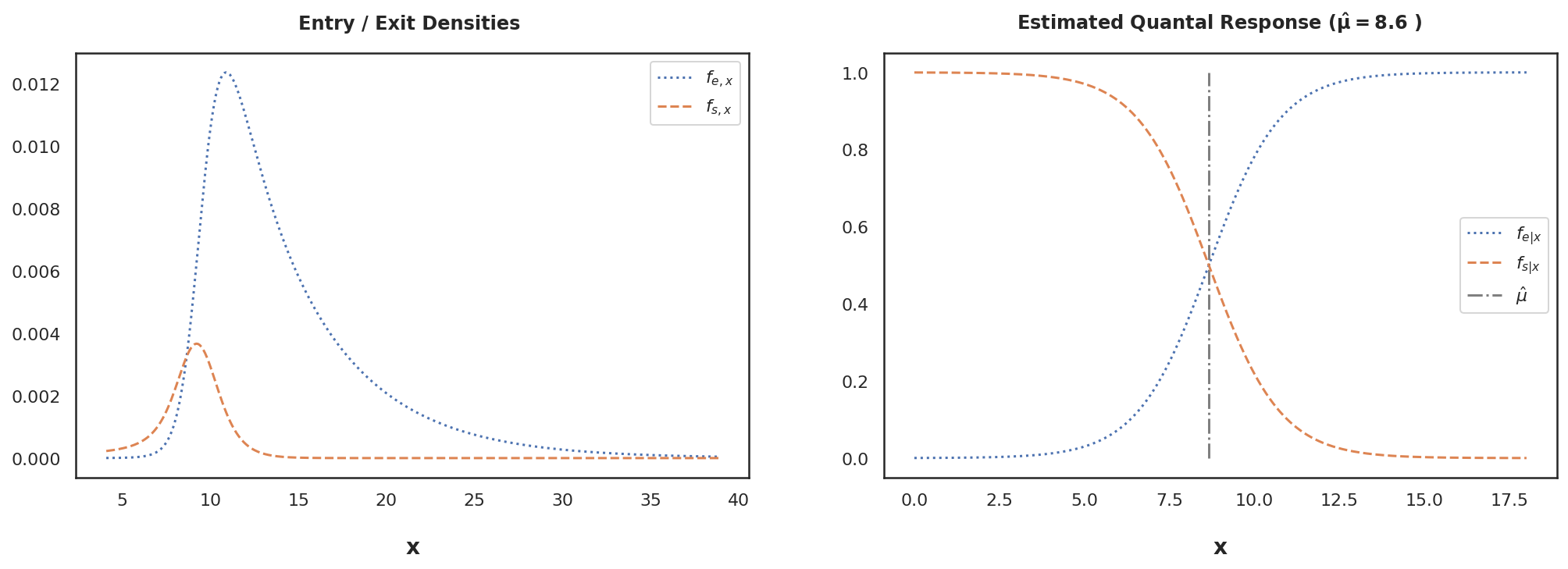}
\caption{\label{acplots}Left: Predictive entry and exit densities $f_{a,x}$. Right: Household Quantal Response Functions $f_{a |x}$. The estimated `tipping point' $\hat{\mu}$ is plotted with a dashed-dotted line.}
\end{figure}

In \ref{qrsefit} we fit the estimated QRSE model to the histograms of
the observed distribution of $x$ for this `full ensemble case', which
covers all US school districts in the 2000-2016 period. In figure
\ref{acplots} we plot the predicted joint action and outcome densities
$f_{a,x}$, alongside the estimated quantal response functions, which
predict the conditional probability of entry and exit of households
into districts given a certain level of educational returns $x$. We
discuss these results in the next subsection.

\clearpage

\section{Discussion}
\label{sec-6}

Our central aim is to study the role that Tiebout-like competition may
play in explaining observed educational returns across school
districts in the US. To do so we implemented a quantal response
statistical equilibrium (QRSE) model, which allowed us to characterize
competition between local jurisdictions as a complex negative feedback
process operating at both the household-level and market-level
scales. The QRSE model used in this paper parsimoniously characterizes
the complex interaction between household jurisdictional choice and
the emergent statistical properties of decentralized education markets
in terms of the parameter vector $\Gamma = [T,S, \mu, \alpha]$. The
observed distribution of educational returns is then explained via the
predictive distribution $\hat{f}_{x; \Gamma}$. In order to better understand the distinct role that both the scale (S
\& T) and location ($\mu$ and $\alpha$) parameters play in explaining
observed patterns, it is useful to plot variations to the individual
parameters holding all others constant. We do so below in figure
\ref{thplots}.

\begin{figure}[htb]
\centering
\includegraphics[width=6in]{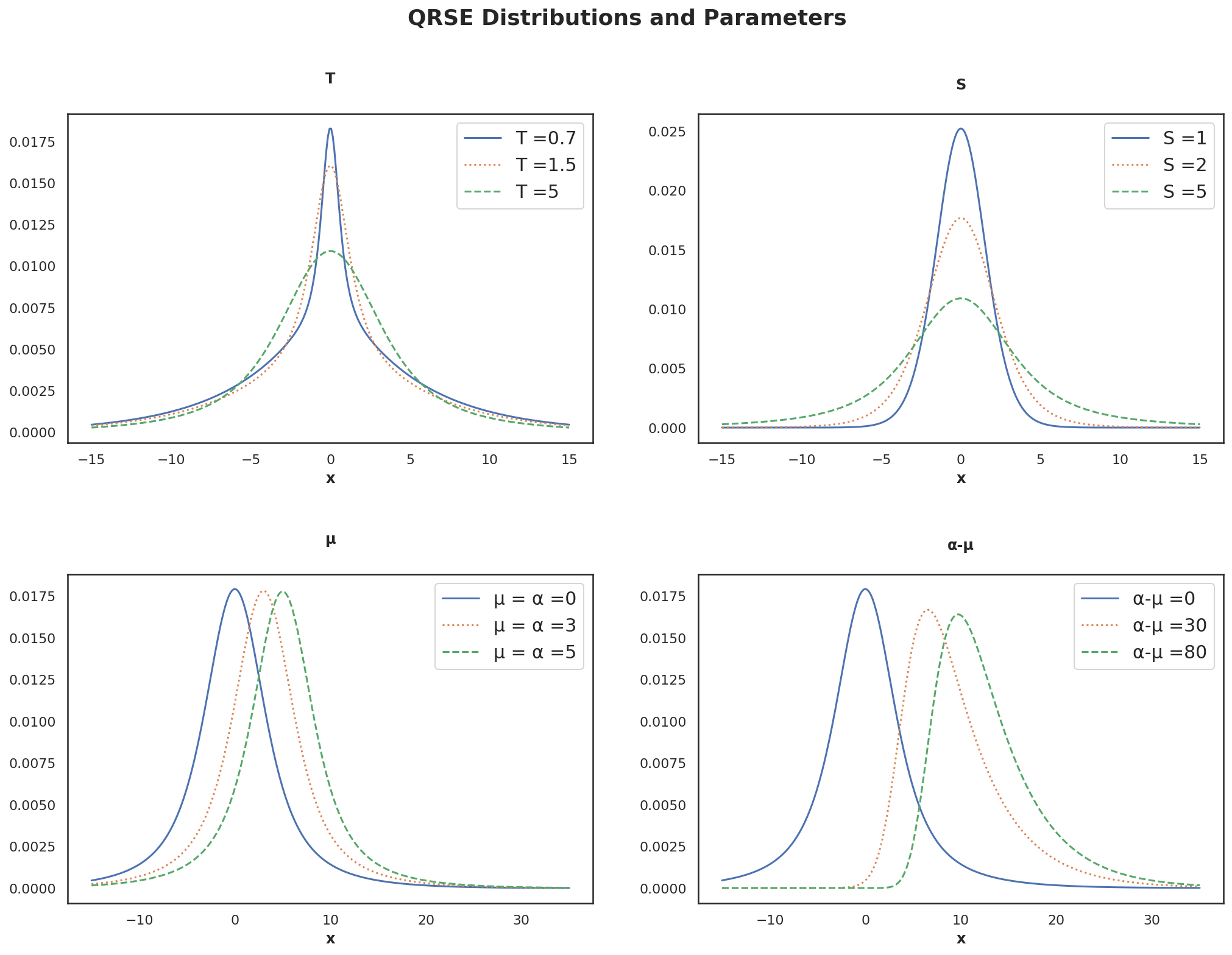}
\caption{\label{thplots}Variations to individual parameter, holding all others constant. The baseline setting is T = 5, S=5, and $\mu$ = $\alpha$ = 0.}
\end{figure}

The $\mu$ and $\alpha$ parameters are particularly relevant in understanding the positive
skewness of the statistical equilibrium distribution that we find for
the full ensemble case. The parameter $\alpha$ estimates a market-level
statistical tendency that acts as the barycenter around which the
household-formed expectation $\mu$ fluctuates. To see how this is
built into the theory, note that in the competitive feedback
constraint in \ref{fbconst}, we write the expectation as $E[x-\alpha]$
and not $E[x]$. In the case where households' expectations of the
local educational returns rate matches the market level tendency, then
$\alpha = \mu$. In that case the distribution is symmetrical and the
estimated values for $\alpha$ and $\mu$ also match the sample mean
$\bar{x}$. Whenever $\mu \neq \alpha$, then the QRSE distribution is
asymmetrical, and positive values for $\alpha - \mu$ in particular
will lend the distribution a more or less sizeable positive skew.
As shown in figure \ref{thplots}, both the behavioral and market scale
parameters T and S predict a lesser/larger concentration of values
around the mode, with lower values lending more peakedness to the
distribution. 

\medskip

The QRSE model explains concentration around modal values as
the consequence of intense competition in decentralized public
education markets. Both relatively purposive households and market
feedbacks work to stabilize educational returns into their statistical
equilibrium distribution. This understanding of competition is consistent with the profit rate
equalization hypothesis that one finds in classical political economy,
and which has been given modern statistical treatments in
\citep{farjoun_laws_2020,schafsim,alfarano_statistical_2012,schafol}.

\medskip 

The statistical equilibrium analysis shows concrete evidence  that there are both
sorting forces and competitive forces at play in determining the equilibrium 
educational returns rate. Tiebout sorting in particular educational markets might 
in fact be signaled by distributions with heavy right tails and positive skew. We believe
this to be the case in the sense that `better sorted' or more
'balanced' subsamples in $x$ will undoubtedly contain a broader set of
tax-service packages that households sort into via the local public
goods and housing markets. In our QRSE model and estimates, the Tiebout `sorting forces' are
captured by the size of the difference $\alpha - \mu$, while the
competitive forces are captured by the size and interaction of the $T$
and $S$ parameters. This leads to future work needing to unpack how
the $\alpha$ parameter is related to median household
income and property values in school districts, given that high income
and property values push public school expenditures far beyond
competitive or modal rates.

Educational returns in school districts across the US for the
2000-2016 period exhibit distinctively peaked, positively skewed
distributions with right tails of variable width. The shaping of their
statistical equilibrium distribution is the outcome of an evolving
process of inter-jurisdictional competition, household residential
sorting on the basis of a broad set of characteristics (such as
income), and shifting policy regimes at the local, state and federal
levels.

\medskip 

Using a statistical equilibrium framework, in this paper we sought to
examine the role played by inter-jurisdictional competition and
household choice in shaping the observed distribution of educational
returns for a full ensemble case that covers all US school districts
in the 2000-2016 period. This is a considerably larger sample than the
ones found in other empirical treatments in the literature, which
usually focus on single states or regions. An
important aspect of our empirical findings is that it corroborates the
need to divorce normative notions about market efficiency from claims about the presence of Tiebout sorting and competition. We proposed a parsimonious model that
meaningfully captures the difference between competitive and sorting
forces via two sets of scale and location parameters. 

\medskip

This empirical analysis also corroborates previous findings in the QRSE
literature regarding the use of feedback constraints as
meaningful characterizations of competition in decentralized market
settings. The histogram and model fit displayed in this paper are
clearly suggestive of the part played by decentralized competition in
sharpening modal peaks, and by elevated (far from competitive) market
rates in creating positive skew.

\newpage 

\clearpage

\bibliographystyle{unsrtnat}
\bibliography{references}  

\end{document}